\documentstyle[11pt,epsf]{article}

\def\be{\begin{equation}}
\def\ee{\end{equation}}
\def\bea{\begin{eqnarray}}
\def\eea{\end{eqnarray}}

\font\extra=msbm10 scaled \magstep1
\def\bbb #1{\hbox{{\extra #1}}}
\begin{document}
\thispagestyle{empty}

\begin{center}
{\LARGE{\bf The supersymmetric modified P\"oschl-Teller \\ [1ex]
and delta--well potentials}}
\vskip1cm

J.I. D\'{\i}az, J. Negro, L.M.~Nieto and  O. Rosas--Ortiz\footnote{On leave of
         absence from {\it Dept. de F\'\i sica}, CINVESTAV-IPN,
         {\it A.P. 14-740, 07000 M\'exico~D.F., Mexico}.}

{\it 
Departamento de F\'\i sica Te\'orica,  Universidad de Valladolid\\
47005 Valladolid, Spain}
\end{center}
\vskip1cm

\begin{abstract} 
\baselineskip=18pt
New supersymmetric partners of the modified P\"{o}schl-Teller and the
Dirac's delta well potentials are constructed in closed form. The
resulting one-parametric potentials are shown to be interrelated by a 
limiting process. The range of values of the parameters for which these
potentials are free of singularities is exactly determined.
The construction of higher order supersymmetric partner potentials is also
investigated.
\end{abstract}

\vglue 4cm 

\begin{description}
\item[Key--Words:] Factorization, modified P\"oschl-Teller potential, 
Dirac delta potential, isospectrality.
\item[PACS:]  03.65.Ge, 03.65.Fd, 03.65.Ca
\end{description}

\newpage

\baselineskip=24pt

\section{Introduction}

The modified P\"{o}schl-Teller potential $V(\alpha,x) = -U_0 (\cosh 
\alpha x)^{-2}$, is one of the few exactly solvable potentials in
quantum mechanics. It was first analyzed by Rosen and Morse
\cite{Ros32}, who found its energy eigenvalues and eigenfunctions.
(This potential $V(\alpha,x)$ is indeed, an hyperbolic version of what
is commonly known as the P\"{o}schl-Teller potential \cite{Pos33}.)
The system is characterized by a finite number of bound states
whose spectrum depends on the parameters $U_0,\alpha>0$, plus a
continuun of scattering states (see in \cite{Alh83} an elegant group
approach); the normalization of the wave functions has been
obtained by Nieto \cite{Nie78}. Barut~{\it et al\/} \cite{Bar87}
studied a  three dimensional version of the problem (with an
additional term proportional to $(\sinh \alpha x)^{-2}$ in the
potential), and using the Infeld-Hull factorization \cite{Inf51} they
constructed ladder operators to determine the bound and scattering
states from the matrix elements of group representations. It is well
known that the modified P\"oschl-Teller potential appears in the
solitary wave solutions of the Korteweg-de Vries equation
\cite{Mat91}, and also that it can be obtained from supersymmetric (susy) 
quantum mechanics \cite{Wit81} as the susy partner of the free
particle potential \cite{Boy88}, being a non trivial example of unbroken 
supersymmetry \cite{Bra85}. 

This potential can be considered
as a type of short range potentials, almost vanishing over most of
their domain, except near zero (where the source resides). The
extreme case of such short range potentials is the  Dirac delta well
$V(x) = -g \delta(x)$,  $g>0$.  Delta potentials have been long used in
field theories, where the main problem arises from the regularization
and renormalization of the values for the physical observables
predicted by the theory itself
\cite{Mit98}. The effects of adding a delta function potential on the 
states of a previously known potential have been computed exactly
by Atkinson \cite{Atk75} for several physical systems. The one
dimensional problem of $N$ particles interacting by means of delta
potentials is one of the simplest many-body problems solved exactly
(see \cite{Ols83} and references quoted therein). As the delta
potential $\delta(x)$ is a distribution rather than a function, it can be
approximated by different families of functions, 
one of them ($V(\alpha, x)$ with an appropriate choice of the
depth~$U_0$) will be investigated in this paper.

Concerning the study of new exactly solvable problems in quantum
mechanics, in the last years there has been remarkable progress 
along different lines: Darboux
transformation \cite{Dar82}, Infeld-Hull factorization \cite{Inf51},
Mielnik factorization \cite{Mie84,Fer84,Suk85}, susy quantum mechanics
\cite{Wit81}, and inverse scattering theory \cite{Zak90}, among
others. It is worth stressing that all of them can be embraced in an
elegant algebraic approach named {\it intertwining technique\/}
\cite{Car79}, which has been successfully applied in the construction
of higher order susy partners \cite{Fer97,Ros98}. 
The usefulness of the intertwining has been also proved 
in the study and interpretation of black-hole perturbations in general
relativity~\cite{And91}. 

The main purpose of this paper is to analyze the susy partners 
associated with the  modified P\"oschl-Teller and Dirac delta
potentials by using the intertwining technique. 
In Section 2, after a short review of results concerning both potentials, 
the properties of the latter are straighforwardly obtained as
a limiting case of the former.
In Section 3 we will determine closed expressions for two new families 
of susy partner potentials of P\"oschl-Teller, paying attention to the 
appearence of singularities (in that case the isospectral properties should be 
considered under the point of view of \cite{Mar98}).
Besides, the limit already mentioned in Section 2 is carefully studied 
here for these two families, obtaining very different behaviour for each 
of them. A direct analysis of the susy delta potential is also carried out to 
check the validity of the previous limits. Finally, in Section 4 other 
varieties of P\"oschl-Teller susy potentials are shown to be easily obtained 
using higher order intertwining.

\section{The essentials of modified P\"oschl-Teller potential}

Let us consider the well known one-dimensional two-parametric 
modified P\"oschl-Teller potential \cite{Lan65}, written in the
following equivalent forms:
\begin{equation}
V(\alpha, x)=-\frac{U_0}{\cosh^2 \alpha x}=
-\frac{\hbar^2}{2m}\ \alpha^2\ \frac{\lambda
(\lambda-1)}{\cosh^2 \alpha x}=-\frac{g \alpha}{2 \cosh^2 \alpha x},
\quad\alpha>0,
\label{pt1param}
\end{equation}
Henceforth, in all of the three versions, the following conditions are
imposed in order to have an attractive potential: $U_0>0, \lambda>1$ or
$g>0$. We also take for simplicity $\hbar^2/2m=1$, and from the last
equality the parameters $\alpha, \lambda$ and $g$ are related by
\begin{equation}
\lambda=\frac12 \left( 1+\sqrt{1+\frac{2g}{\alpha}} \right)>1. 
\label{lambda}
\end{equation}
The bound states ($E<0$) for this potential can be obtained by using the
traditional recipe of transforming the stationary Schr\"odinger equation
into an hypergeometric equation, with parameters
\begin{equation}
a = \frac12 \left( \lambda
-\frac{\sqrt{|E|}}{\alpha}\right) ,\qquad 
b = \frac12 \left( \lambda
+\frac{\sqrt{|E|}}{\alpha}\right) ,\qquad 
c = \frac12,
\label{abc}
\end{equation}
The  general solution is
\begin{eqnarray}
\psi(x)  =  (\cosh\alpha x)^{\lambda} \!\!&\!\! \!\!&\!\!  \left[   A\
{}_2F_1(a,b;1/2;-\sinh^2\alpha x) \right. \nonumber \\ 
 \!\!&\!\! \!\!&\!\!   \left. +B\ (\sinh\alpha x) \, 
{}_2F_1(a+1/2,b+1/2;3/2;-\sinh^2\alpha x) \right]. 
\label{solgen}
\end{eqnarray}
The normalization condition on these eigenfunctions allows to 
determine the energy spectrum, which is found to be
\begin{equation}
E_n= - \alpha^2\, (\lambda-1-n)^2,\quad n\in{\bbb N},\ 0\leq
n<\lambda-1,
\label{spectrum}
\end{equation}
or just in terms of $\alpha$ and $g$:
\begin{equation}
E_n=-\alpha^2\,  \left( \frac12
\sqrt{1+\frac{2g}{\alpha}} -\frac12 -n \right)^2,\quad
n=0,1,2,\dots< \frac12 \sqrt{1+\frac{2g}{\alpha}} -\frac12.
\label{enpt}
\end{equation}
From (\ref{spectrum}) and $\lambda >1$, the
energy for $n=0$ always belongs to the spectrum
of $V(\alpha,x)$. Calling $ N$ the biggest possible value of $n$ in 
(\ref{enpt}), the total number of bound states is ${ N}+1$. 
From there, the following relationship between the parameters
$\alpha$ and $ N$ holds
$$
\frac{2g}{(2{ N}+3)^2-1}\leq  \alpha <
\frac{2g}{(2{ N}+1)^2-1}.
$$
If ${ N}=0$, then $\alpha\geq g/4$, and there is just one bound state
such that it has the lowest energy $E_0$, and for which we have 
\begin{equation}
\label{ground}
a_0= c_0=1/2, \ \ b_0= \lambda - 1/2,\ \ A\neq 0, \ \ B=0, \ \  E_0= -
\alpha^2\, (\lambda -1)^2. 
\end{equation}
The corresponding normalized wave function turns out to be
\begin{equation}
\psi_0(x)  = C_0  (\cosh \alpha x)^{\lambda}\, {}_2F_1\!\left( \frac12,
\lambda - \frac12 ; \frac12 ;-\sinh^2\alpha x \right) 
=\sqrt{ \frac{\alpha \,\Gamma ( \lambda-1/2)}{\sqrt{\pi} \,
\Gamma(\lambda-1)}}
(\cosh\alpha x)^{1 - \lambda}.
\label{groundfunc}
\end{equation}
It is also interesting to remark that (\ref{pt1param}) is a {\it
transparent\/} potential ({\it i.e.\/} the reflexion
coefficient is equal to zero) when the following condition is verified
\cite{Lan65}: 
\begin{equation}
{g}={2\alpha}k(k+1),\quad k=0,1,2,\dots
\label{transparent1}
\end{equation}
Using our notation this implies
\begin{equation}
\lambda = k+1, \quad k=1,2,\dots
\label{transparent2}
\end{equation}
When $k=0$ we have the case of the free particle, which was
excluded from the very begining.

The last form of the potential $V(\alpha, x)$ given in  (\ref{pt1param}) can
be directly related to the Dirac delta potential $V_D(x)=-g\delta(x)$ just by
taking the limit $\alpha\to\infty$:
\begin{equation}
\lim_{\alpha\to\infty} V(\alpha,x)= -g\, \delta(x).
\label{limiteadelta}
\end{equation}
As it is well known, this $\delta$-well has a unique bound state with
eigenvalue
\begin{equation}
E_\delta =-\left({g}/{2}\right)^2 , 
\label{deltalevel}
\end{equation}
and normalized wave function
\begin{equation}
\psi_\delta(x)= \sqrt{{g}/{2}}\
e^{- {g}|x|/{2}}.
\label{deltafunc}
\end{equation}
These results can be obtained directly from the modified P\"{o}schl-Teller
potential by proving that the limiting relationship (\ref{limiteadelta})
between both types of potentials is also inherited by their eigenfunctions
and energy eigenvalues. Indeed, from the the P\"oschl-Teller ground state
energy level (\ref{ground}) we have
\begin{equation}
\lim_{\alpha\to\infty} E_0= - 
\lim_{\alpha\to\infty} \frac{\alpha^2}4\left(
\frac{g}{\alpha}- \frac{g^2}{2\alpha^2}+\cdots
\right)^2=-\left(\frac{g}{2}\right)^2=E_\delta .
\end{equation}
A similar analysis can be done for the ground state eigenfunction
(\ref{groundfunc}). Observe that $\psi_0(0)=C_0$; if $x\neq 0$, then the
limit $\alpha\to\infty$ is equivalent to $\alpha x\to\pm\infty$, hence it
can be calculated from the asymptotic form of the function. To find it, we
use
\begin{equation}
\cosh\alpha x\sim \frac{e^{\alpha |x|}}2,\qquad
\sinh \alpha x\sim\pm \frac{e^{\alpha |x|}}2,\qquad (\alpha>0),
\label{asymhyperbol}
\end{equation}
and also the fact that, for $z\to-\infty$, we have \cite{Erd53}
\begin{equation}
{}_2F_1(a,b;c;z)\sim \Gamma(c)\left\{
\frac{\Gamma(b-a)}{\Gamma(b)\Gamma(c-a)}(-z)^{-a} +
\frac{\Gamma(a-b)}{\Gamma(a)\Gamma(c-b)}(-z)^{-b}  \right\},\ \
|{\rm arg}(-z)|<\pi.  
\label{asymhypergeom}
\end{equation}
Then, from
(\ref{ground}) and (\ref{asymhypergeom}), we have for $|z|\to\infty$
\begin{equation}
{}_2F_1\!\left( \frac12,b_0;\frac12;-z^2 \right) \sim z^{-2b_0}.
\end{equation}
In addition we have the following asymptotic behaviour for
$\alpha\to\infty$: $2b_0\sim 1 + {g}/{\alpha}$. Hence
\begin{equation}
\lim_{\alpha\to\infty} \psi_0(x)  = C_\infty \lim_{\alpha\to\infty}
2^{b_0-\frac12}\ e^{-|x|\alpha(b_0-\frac12)}=C_\infty \ e^{- {g}|x|/{2}},
\label{estadofundamental}
\end{equation}
where $C_0\to C_\infty$.
From (\ref{groundfunc}) it is clear that $C_\infty=\sqrt{g/2}$, and then
equation (\ref{estadofundamental})  states the connection between
(\ref{groundfunc}) and (\ref{deltafunc}) when the limiting relationship
(\ref{limiteadelta}) is satisfied.

Finally, it is also well known that the $\delta$-well potential is
transparent. This property is also deduced from the P\"{o}schl-Teller
potential by taking the limit $\alpha\to\infty$ in (\ref{transparent2}),
and using (\ref{lambda}), which gives the solution $k=0$. 

\section{Susy modified P\"oschl-Teller potential}

Let us consider now the problem of finding the
supersymmetric partner of the modified P\"oschl-Teller potential
$V(\alpha,x)$. We look for a first order differential operator $A=\frac{d}{dx}
+\beta(x)$ and a partner potential ${\widetilde V}(\alpha,x)$ such that 
the following interwining relationship holds:
\begin{equation}
\left[- \frac{d^2}{dx^2} + {\widetilde V}(\alpha,x)  \right]\, A=
A\, \left[ -  \frac{d^2}{dx^2} + V(\alpha,x) \right].
\label{intertw}
\end{equation}
The new potential ${\widetilde V}(\alpha,x)$ is
related to $V(\alpha,x)$ through the following susy relationship
\begin{equation}
{\widetilde V}(\alpha,x)=V(\alpha,x)+  2\beta'(x),
\label{firstvtilde}
\end{equation}
where $\beta(x)$ is a solution of the Riccati equation
\begin{equation}
\beta^2(x) -\beta'(x)= V(\alpha,x)-\epsilon,
\label{riccpt}
\end{equation}
with $\epsilon$ an integration constant, which turns out to be the
factorization energy. There is an immediate particular solution of equation
(\ref{riccpt}) in the form of an hyperbolic tangent,
$\beta_{0}= D\, \tanh \alpha x$, with $D$ depending on $\alpha$. The
introduction of $\beta_{0}$ in (\ref{riccpt}) gives
\begin{equation}
D^{+} = -\alpha\lambda,\qquad 
D^{-} = -\alpha(1-\lambda).
\label{Dpm}
\end{equation}
Therefore, we have two different particular solutions of (\ref{riccpt})
\begin{equation}
\beta_{0}^{\pm}(\alpha,x) = D^{\pm} \tanh \alpha x,
\label{betaparticular}
\end{equation}
associated with two different factorization energies
\begin{equation}
\epsilon^\pm=-(D^{\pm})^2 = -\frac{\alpha^2}{2} \left( 1 + \frac{g}{\alpha}
\pm \sqrt{ 1 + \frac{2g} {\alpha}} \right).
\label{epsilonpm}
\end{equation}
Remark that these factorization energies can be formally identified with two 
values of the spectrum formula (\ref{enpt}):
\be
\epsilon^-=E_0,\qquad  \epsilon^+= E_{-1}.
\label{epsilones}
\ee
Then, the general solutions of the
Riccati equation (\ref{riccpt}) for the above factorization energies can
be found to be
\begin{eqnarray}
\beta^{+}_\zeta (\alpha,x)  \!\!&\!\! =\!\!&\!\!   D^+\, \tanh \alpha x -
\frac{d}{dx}\, \ln \left( 1- \zeta \int^x (\cosh \alpha y)^{{2D^+}/\alpha}\,
dy\right), \label{betaplusminus}\\
\beta^{-}_\xi (\alpha,x)   \!\!&\!\! =\!\!&\!\!    D^-\, \tanh \alpha x -
\frac{d}{dx}\, \ln \left( 1- \xi \int^x (\cosh \alpha y)^{{2D^-}/\alpha}\,
dy\right), \label{betaplusminus2}
\end{eqnarray}
where $\zeta, \xi$ are two new independent integration constants; 
when they
are taken to be zero, we recover the particular solutions $\beta_0^\pm
(x)$. It must be clear that, in fact,  we have obtained
two different families of intertwining operators
\begin{equation}
A^+_\zeta =\frac{d}{dx} +\beta^+_\zeta (x), \qquad
A^-_\xi =\frac{d}{dx} +\beta^-_\xi (x),
\label{Apm}
\end{equation}
generating two different families of susy partners (\ref{firstvtilde}) of
the potential (\ref{pt1param}). 
Up to now, it has been usual to consider only the susy partners of a given
potential constructed by taking particular solutions of the Riccati
equation \cite{Wit81,Boy88}. For the potential we are dealing with, the 
cases $\zeta=0$ or $\xi=0$ give interesting results, and will be obtained 
just as byproducts of (\ref{betaplusminus})--(\ref{betaplusminus2}). 
In principle, it is possible to find solutions associated to other factorization 
energies, but they will produce very awkward expressions, without adding 
new relevant information to the problem we are dealing with.

Let us analyze now the results coming out when the general solutions
(\ref{betaplusminus})--(\ref{betaplusminus2}) are taken into account.
Observe that the integrals appearing there can be expressed in a closed form
as
\begin{equation}
\int^x (\cosh \alpha y)^q\, dy= -\frac{2^{-q}\,
e^{-\alpha q x}}{\alpha q}\ {}_2F_1 \left(-\frac{q}2,-q;
1-\frac{q}2;-e^{2\alpha x}\right)+{\rm constant}.
\label{laintegral}
\end{equation}
This expression will be used next to construct the closed form of the
susy partner potentials.

\subsection{The two parametric family of potentials  ${\widetilde
V}_\zeta^+ (\alpha,x)$}

In this case, taking into account  (\ref{Dpm})  and  (\ref{betaplusminus}),
we see that the exponent in (\ref{laintegral}) is negative, indeed
$q=-2\lambda$. The definite integral exist in the whole real axis, and 
in this case we can define the function
\begin{equation}
M(\lambda,\alpha,x)=\int^x_{0} (\cosh \alpha y)^{-2\lambda}\, dy=
\frac{2^{2\lambda}\, e^{2\alpha \lambda x}}{2\alpha \lambda}\ {}_2F_1
\left(\lambda,2\lambda; 1+\lambda;-e^{2\alpha x}\right)-
\frac{\sqrt{\pi}\,\Gamma(\lambda)}{2\alpha\,\Gamma(\lambda+1/2)} .
\label{29}
\end{equation}
A typical plot of $M(\lambda,\alpha,x)$ is shown in Figure~1 for general
values of the parameters $\alpha$ and $\lambda$.  It is quite clear that
this function is odd in the variable $x$, and it is monotonically
increasing from its minimum value $M(\lambda,\alpha,-\infty)= -
{\sqrt{\pi}\,\Gamma(\lambda)}/(2\alpha\,\Gamma(\lambda+1/2))$ to its
maximum value $M(\lambda,\alpha,+\infty)= |M(\lambda,\alpha,-\infty)|$. It
is interesting to remark the resemblance between the form of this function
and that of the error function ${\rm erf\,} (x)=\frac2{\sqrt{\pi}}\,
\int_0^x e^{-y^2}\, dy$. We will immediately see that, indeed, for the
P\"oschl-Teller problem, $M(\lambda,\alpha,x)$ plays a role completely
analogous to that played by the error function when determining the
one-parametric family of susy partners of the harmonic oscillator
potential  \cite{Mie84}. 

If we introduce now the function
\begin{equation}
\Omega^+_\zeta(\lambda,\alpha,x): =
\frac{(\cosh \alpha x)^{-2\lambda}}{1-
\zeta\, M(\lambda,\alpha,x) }= -\frac1{\zeta}\frac{d}{dx}\ln \left[  1-
\zeta\, M(\lambda,\alpha,x) \right]
\end{equation}
we have for
$\beta^+$ in (\ref{betaplusminus}) the following expression:
\begin{equation}
\beta^+_\zeta (\alpha,x)= -\alpha\lambda\, \tanh \alpha x +\zeta\,
\Omega^+_\zeta(\lambda,\alpha,x) =
\frac{d}{dx} \ln \left[ (\cosh\alpha x)^{\lambda}\
\Omega^+_\zeta(\lambda,\alpha,x) \right].
\label{betapluszeta}
\end{equation}
From here, and using (\ref{firstvtilde}), we can evaluate the associated 
susy partner potential, which  turns out to be
\begin{eqnarray}
\nonumber
{\widetilde V}_\zeta^+ (\alpha,x)  \!\!&\!\! =\!\!&\!\!  
-\alpha^2\left(1+\frac{g}{2\alpha}+
\sqrt{1+\frac{2g}{\alpha}} \right)  \frac1{\cosh^2\alpha x} \\ [1ex]
 \!\!&\!\!  \!\!&\!\!  
- 4\lambda\alpha\zeta\,
\Omega^+_\zeta(\lambda,\alpha,x)\, \tanh
\alpha x+2 (\zeta\,
\Omega^+_\zeta(\lambda,\alpha,x))^2.
\label{vtildepluszeta}
\end{eqnarray}
It is obvious that the singularities of ${\widetilde V}_\zeta^+$
correspond to the singular points of the function
$\Omega^+_\zeta(\lambda,\alpha,x)$. It can be proved that this function is
free of singularities in the following range of values of $\zeta$
\begin{equation}
|\zeta| < \frac1{M(\lambda,\alpha,+\infty)} =
\frac{2\alpha\,\Gamma(\lambda+1/2)}{\sqrt{\pi}\,\Gamma(\lambda)} .
\label{dominioplus}
\end{equation}
As we pointed out before,  if  we compare our results with
the pioneering Mielnik's work on the harmonic oscillator \cite{Mie84}, one
can appreciate that the roles played there by the error function and his
parameter $\gamma$, are performed here by $M(\lambda,\alpha,x)$ and
the inverse of $\zeta$. The characteristic features of
$M(\lambda,\alpha,x)$ determine the existence of susy partner potentials
wich are free of singularities, and are therefore almost isospectral to the
modified P\"oschl-Teller potential. (A similar analysis can be done for an
equivalent integral appearing in  \cite{Fer84}.)

Let us work now in this range of values of the parameter $\zeta$. The
potential (\ref{vtildepluszeta}) corresponds to the following family of 
almost isospectral Hamiltonians
\begin{equation}
{\widetilde H}^+_\zeta  := - \frac{d^2}{dx^2} + {\widetilde V}^+_\zeta
(\alpha,x)  = A^+_\zeta  \,(A^+_\zeta )^\dagger +\epsilon^+.
\label{widetildeh}
\end{equation}
It is well known that the eigenfunctions of ${\widetilde
H}^+_\zeta $ can be constructed by acting with the operator $A^+_\zeta$ of 
(\ref{Apm}) on  the eigenfunctions $\psi_n$ of $H$, which is factorized as
\begin{equation}
H=-\frac{d^2}{dx^2}+V (\alpha,x) = (A^+_\zeta )^\dagger \, A^+_\zeta  +
\epsilon^+,
\label{hfactorizado}
\end{equation}
and are given by ${\widetilde \psi}^+_n(\zeta,x)\propto A^+_\zeta 
\,\psi_n(x)$,   provided that $A^+_\zeta \,\psi_n(x)\neq 0$ and
${\widetilde
\psi}^+_n(\zeta,x) \in L^2({\bbb R})$. There is also the possibility of an
extra eigenfunction ${\widetilde \varphi}^+ (\zeta,x)$ satisfying
\begin{equation}
(A^+_\zeta )^\dagger{\widetilde \varphi}^+ (\zeta,x)=0,
\label{missing}
\end{equation}
which will be called ``missing state''.

The first point we want to stress is that according to (\ref{epsilones})
$\epsilon^+=E_{-1}$, wich is an energy level not allowed in the spectrum
of the initial Hamiltonian $H$. Hence, from (\ref{hfactorizado}) it is
clear that there is no eigenfunction ${\widehat \psi}$ of $H$ annihilated
by $A^+_\zeta$. Therefore, the eigenfunctions of ${\widetilde H}^+_\zeta $
are given by the normalized functions
\begin{equation}
{\widetilde \psi}^+_n (\zeta,x)= (E_n-\epsilon^+)^{-1/2}A^+_\zeta \psi_n(x),
\quad n=0,1,\dots,
\end{equation}
plus the missing state solving (\ref{missing}), which properly normalized
reads
\begin{equation}
{\widetilde \varphi}^+(\zeta, x)=
\sqrt{
\frac{1-\zeta^2\
M^2(\lambda,\alpha,+\infty)}{2\, M(\lambda,\alpha,+\infty)}}\ 
(\cosh \alpha x)^\lambda\
\Omega^+_\zeta(\lambda,\alpha,x).
\label{missingstate}
\end{equation}
Remark that the non-singularity condition (\ref{dominioplus}) appears  
here again, although in this case it is required for the missing state to be
normalizable.
Note also from (\ref{widetildeh}) that ${\widetilde \varphi}^+ (\zeta,x)$ is
clearly the eigenfunction of
${\widetilde H}^+_\zeta $ with eigenvalue $\epsilon^+$. This is the reason
why we named it missing state. The spectrum of ${\widetilde H}^+_\zeta $ is
given by the set   
$\{ E_n; n=0,1,\dots \}$ plus a new level at $\epsilon^+=E_{-1}$. The
conclusion is immediate: the family of potentials ${\widetilde V}^+_\zeta
(\alpha,x)$ is not strictly isospectral to its susy partner $V(\alpha, x)$: it 
has the same levels plus an additional one which is placed below all of
them. Let us remark that, due to the annihilation of the missing state
${\widetilde
\varphi}^+ (\zeta,x)$ in (\ref{missingstate}) by the intertwiner 
$A^+_\zeta$, the missing state has no susy partner, and the couple of
almost isospectral  Hamiltonians $H$ and $H^+_\zeta$ corresponds to a case
of unbroken susy.

In Figure~2 we plotted the asymmetric double well corresponding to
the susy partner potential ${\widetilde V}^+_\zeta (\alpha,x)$ given by
(\ref{vtildepluszeta}), with $\alpha=0.1$, $\lambda=3$, and
$\zeta=0.0937$. The three bound states of this potential are also
represented with dotted horizontal lines. Note that the potentials
${\widetilde V}^+_\zeta (\alpha,x)$ present features that makes
them interesting for physical applications:
(i)  we are able to know their spectra in an exact form, and
by adjusting the parameters we
can have the desired number of bound states, and (ii) the shape of the 
potential can be modified to allow interesting tunnelling effects.

Let us consider now the limit $\alpha\to\infty$ of the potentials
(\ref{vtildepluszeta}). 
There are two terms containing the
function $\Omega^+_\zeta(\lambda,\alpha,x)$; using (\ref{asymhyperbol}),
(\ref{asymhypergeom}), and (\ref{dominioplus}), it can be easily proved
that it has the following behaviour for large values of $\alpha$:
\begin{equation}
\Omega^+_\zeta(\lambda,\alpha,x)
{}_{\widetilde{\alpha\to\infty}} 
\left\{   
\begin{array}{cc}
4 e^{-2\alpha|x|}, & x\neq 0, \\
1, & x=0,
\end{array}
\right. 
\end{equation}
giving a discontinuous function in the limit, but which is zero almost
everywhere. In addition, the product $\alpha\,
\Omega^+_\zeta(\lambda,\alpha,x)\to 4\delta(x)$, and $\alpha\,
\Omega^+_\zeta(\lambda,\alpha,x) \tanh \alpha x\to 0$. Hence, the only
relevant part in this potential would be that coming from the first term. 
But it diverges very badly as $-4 \alpha \delta(x)$, and therefore we do not end with
a physically interesting potential. 

\subsection{The two parametric family of potentials  ${\widetilde
V}_\xi^- (\alpha,x)$}

The study of the other family of potentials coming out from
(\ref{betaplusminus2}) can be done according to the lines already followed
in the previous subsection. Nevertheless, there are important differences
between the results obtained in both cases. First of all, from equations
(\ref{Dpm})  and (\ref{betaplusminus2}), the exponent in
(\ref{laintegral})  is now positive: $q=2(\lambda-1)$. As a consequence,
if we try to evaluate the integral in the whole real axis we get a
divergent result.  Nevertheless, it is useful to introduce a function
similar to $M(\lambda,\alpha,x)$, let us call it
\begin{eqnarray}
L(\lambda,\alpha,x)   \!\!&\!\! =\!\!&\!\!   \int^x_{0} (\cosh \alpha
y)^{2(\lambda-1)}\, dy 
\label{L} \\
 \!\!&\!\! =\!\!&\!\!    -\frac{e^{-2\alpha (\lambda-1) x}}{2^{2(\lambda-1)}\,
2\alpha (\lambda-1)}\  {}_2F_1\left(1-\lambda,2-2\lambda;
2-\lambda;-e^{2\alpha x}\right)+
\frac{\sqrt{\pi}\,\Gamma(2-\lambda)}{2\alpha(\lambda-1)\,
\Gamma(\frac32-\lambda)}. \nonumber 
\end{eqnarray}
This function is also odd and takes arbitrary positive values for $x>0$ and
arbitrary negative values for $x<0$. Using the asymptotic behaviour of the
hypergeometric functions (\ref{asymhypergeom}), it is  very easy to prove
that the limit $\alpha\to\infty$ of (\ref{L}) is
\begin{equation}
\lim_{\alpha\to\infty} L(\lambda,\alpha,x)=
 \left( \frac{e^{g|x|}-1}{g} \right) {\rm sgn\,}x,
\label{limitl}
\end{equation}
where ${\rm sgn\,}x$ denotes the function sign of $x$, and we have used
the fact that $\alpha$, $\lambda$ and $g$ are related through equation
(\ref{lambda}). We will use $L(\lambda,\alpha,x)$ to define the function
\begin{equation}
\Omega^-_\xi(\lambda,\alpha,x): =
\frac{(\cosh \alpha x)^{2(\lambda-1)}}{1-
\xi\, L(\lambda,\alpha,x) }= -\frac1{\xi}\frac{d}{dx}\ln \left[  1-
\xi\, L(\lambda,\alpha,x) \right],
\label{omegaminus}
\end{equation}
from which the following expression for $\beta^-_\xi$ in
(\ref{betaplusminus2}) is obtained
\begin{equation}
\beta^-_\xi (\alpha,x)= \alpha(\lambda-1) \tanh \alpha x +\xi\
\Omega^-_\xi(\lambda,\alpha,x) =
\frac{d}{dx}\ln \left[ (\cosh\alpha x)^{1-\lambda}\
\Omega^-_\xi(\lambda,\alpha,x) \right].
\label{betaminusxi}
\end{equation}
Using this expression and (\ref{firstvtilde}) we compute the new
susy partner potential
\begin{eqnarray}
\nonumber
{\widetilde V}_\xi^- (\alpha,x)  \!\!&\!\! =\!\!&\!\!  
-\alpha^2\left(1+\frac{g}{2\alpha}-
\sqrt{1+\frac{2g}{\alpha}} \right)  \frac1{\cosh^2\alpha x} \\ [1ex]
 \!\!&\!\! \!\!&\!\!  
+4\alpha(\lambda-1)\xi\
\Omega^-_\xi(\lambda,\alpha,x)\, \tanh
\alpha x  +2 (\xi\ \Omega^-_\xi(\lambda,\alpha,x))^2.
\label{vtildeminuszeta}
\end{eqnarray}
Due to the behaviour of $L(\lambda,\alpha,x)$, it is quite clear that for
any choice of $\xi\neq 0$ the function $\Omega^-_\xi(\lambda,\alpha,x)$
presents a singular point, and therefore the potential ${\widetilde
V}_\xi^- (\alpha,x)$ is always singular, in contradistinction to the case
of ${\widetilde V}_\zeta^+ (\alpha,x)$ considered before. The presence of
the singularity suggest that the results could be interpreted according to
the method developed in \cite{Mar98}: the susy partner potentials are not
directly related by isospectrality to the original potential
$V(\alpha,x)$, but to a different problem consisting of this modified
P\"oschl-Teller potential plus an infinite barrier potential placed
precisely at the position where ${\widetilde V}_\xi^- (\alpha,x)$ has its
singular point. 

The case $\xi= 0$ is  interesting enough to be considered separately.
It gives just the particular solution
${\widetilde V}^-_0 (\alpha,x)$, which is free of singularities.
The susy partner potentials~(\ref{vtildeminuszeta}) correspond to a
factorization energy $\epsilon^-$, which according to  (\ref{epsilones}) 
 is $\epsilon^-=E_0$. Therefore, there is an eigenstate
$\widehat{\psi}$ of $H$ which is annihilated by the intertwining operator
$A^-_0\, \widehat{\psi} (x)=0$, and we can write 
$$
\widehat{\psi} (x)\propto \exp\left[ -\int^x \beta_0^-(y)\, dy \right]= 
(\cosh \alpha x)^{1-\lambda},
$$
which is precisely the square integrable function $\psi_0(x)$ given in
(\ref{groundfunc}), the ground state of $H$. The eigenfunctions of
${\widetilde H}^-_0 $ are then given by
\begin{equation}
{\widetilde \psi}^-_n (x)= (E_n-\epsilon^-)^{-1/2}A^-_0  \psi_n(x).
\quad n=1,2,\dots,
\end{equation}
In the present case the possible missing state solving (\ref{missing}) would
be
$$
{\widetilde \varphi}^- (x)\propto \exp\left[ \int^x \beta_0^-(y)\, dy 
\right]=  (\cosh \alpha x)^{\lambda-1}.
$$
As $\lambda>1$, this function is not square integrable, and therefore it  
has not physical meaning as an eigenfunction of ${\widetilde H}^-_0 $ with
eigenvalue $\epsilon^-$. The spectrum of ${\widetilde H}^-_0 $ is given  
simply by $\{ E_n; n=1,2,\dots \}$. Remark that, like in the previous case,
this new Hamiltonian is not either strictly isospectral to $H$, although the
reason is just the opposite: now the susy process eliminates one state of
$H$ without creating a new one which can substitute it, while in the
previous situation a new state was created, but keeping the initial spectrum.
Hence, we have another example of unbroken susy encoded in the spectrum
of the couple $H$ and ${\widetilde H}^-_0$.

It is interesting to evaluate the limit of the previous results for
$\alpha\to\infty$,  in order to do that, we first analize the asymptotic
behaviour of the factorization energy (\ref{epsilonpm}):
\begin{equation}
\epsilon^-=E_0\sim -\left(\frac{g}2\right)^2+\frac{g^3}{4\alpha}+
O\left(\frac1\alpha\right)^2.
\end{equation}
For $0< \alpha<g/4$ the susy potential ${\widetilde V}^-_0 (\alpha,x)$ in
(\ref{vtildeminuszeta}) is always attractive; on the other hand, when
$\alpha>g/4$ the potential becomes always repulsive;  finally, when
$\alpha=g/4$ the potential vanishes identically. This change on the
character atractive or repulsive of ${\widetilde V}^-_0 (\alpha,x)$
according to the values of $\alpha$ has a {\it strong\/} physical meaning.
It is related to the fact that in the susy process, the ground state level
$E_0$ is eliminated from the spectrum of ${\widetilde H}^-_0 $, being
always a member of the spectrum of $H$, irrespectively of the value of
$\alpha$. In particular, when $\alpha>g/4$ the potential $V(\alpha,x)$ has
only this bound state (see coment after equation (\ref{enpt})), and
therefore, in the same interval the new potential ${\widetilde V}^-_0
(\alpha,x)$ has not bound state at all. 

In Figure~3 we plotted three members of the original family
(\ref{pt1param}) of modified P\"oschl-Teller potentials $V(\alpha,x)$  (the
three thicker curves), and also, in the same type of lines, but thinner, their
corresponding susy partner potentials ${\widetilde V}^-_0  (\alpha,x)$ given
by (\ref{vtildeminuszeta}). The values of the  parameters are indicated in
the caption. Remark that although the initial modified P\"oschl-Teller
potentials are always negative, their susy potentials are less negative (in the
case of the dotted line) or become even positive (as in the cases of dashed
and solid curves).

An important detail to be stressed is that in the limit
$\alpha\to\infty$ the potential ${\widetilde V}^-_0  (\alpha,x)$ of
(\ref{vtildeminuszeta}) has a well defined behaviour (unlike the situation for
${\widetilde V}^+_0 (\alpha,x)$): it becomes the delta barrier
\begin{equation}
\lim_{\alpha\to\infty} {\widetilde V}^-_0  (\alpha,x)= + g\,
\delta(x),\qquad  g>0.
\label{deltaparriba}
\end{equation}
It is quite remarkable the difference with the limit of the initial
potential $V(\alpha,x)$, which was a delta well, as
it has been shown in equation (\ref{limiteadelta}). 

Another interesting point to be considered is the analysis of the limit
$\alpha\to\infty$ of (\ref{vtildeminuszeta}), which can be evaluated even
for the case $\xi\neq 0$. We already have all the information needed to
write down this result, indeed: $\alpha(\lambda-1)\to g/2$, $\tanh\alpha
x\to {\rm sgn\,}x$, plus equations (\ref{deltaparriba}), (\ref{limitl}),
and (\ref{omegaminus}). We get the following: 
\begin{eqnarray}
\lim_{\alpha\to\infty} \beta_\xi^- (\alpha,x)   \!\!&\!\! =\!\!&\!\!  
\frac{g}2\,  {\rm sgn\,} x+\xi\ \frac{ e^{g|x|}}{1-\xi
\left(
\frac{e^{g|x|}-1}{g} \right)\, {\rm sgn\,}x}  ,
\label{limitbetaminustodelta}
\\ [2ex]
\lim_{\alpha\to\infty} {\widetilde V}_\xi^- (\alpha,x)
 \!\!&\!\! =\!\!&\!\!    g\, \delta(x)+
\frac{2 g\,\xi\ e^{g|x|}\,  {\rm sgn\,} x}{1-\xi \left(
\frac{e^{g|x|}-1}{g}
\right) \, {\rm sgn\,}x }+\frac{2\xi^2\ e^{2g|x|}}{\left[1-\xi
\left( \frac{e^{g|x|}-1}{g}
\right) \, {\rm sgn\,}x\right]^2}.
\label{limitvminustodelta}
\end{eqnarray}
Again, we have used the fact that the parameters $\alpha$, $\lambda$
and $g$ are related through (\ref{lambda}). Observe that after the limit
process, we obtain a function with three different discontinuities: 
\begin{enumerate}
\item
The function blows up like $2\, (x-x_s)^{-2}$ at the singular point
$$
x_s=\frac{{\rm sgn\,}\xi}g\, \ln\left( 1+ \frac{g}{|\xi|} \right).
$$
\item 
At the origin,  we get a divergence of the type $+g\, \delta(x)$.
\item
There is a finite jump discontinuity at the origin, due to the presence of
the sign function in (\ref{limitvminustodelta}). This discontinuity is
concealed by the presence of the superposed Dirac delta centred also at
$x=0$.
\end{enumerate}
These remaks are clearly illustrated  on Figure~4, where we have
plotted the limit case (without the delta distribution at the origin), plus one
intermediate case. The dotted curve represents the plot of one of the susy
partners of the modified P\"oschl-Teller potential (\ref{vtildeminuszeta}), for
the following values of the parameters:
$\alpha=1.9$, $\lambda=1.216$ (or equivalently $g=1$), $\xi=-0.05$. The
solid curve represents the susy Dirac delta   potential for $g=1$ and
$\xi=-0.05$, and it is a limiting case of the dotted curve when
$\alpha\to\infty$ (the delta contribution comes out from the dotted hump).

\subsection{The connection with the susy Dirac delta potential}

Let us consider now the intertwining relationship (\ref{intertw}) for the
Hamiltonian associated with the Dirac delta well potential
$V_D(x)=-g\delta(x)$. Equations (\ref{firstvtilde}) and (\ref{riccpt})
also hold, with $V_D(x)$ as the known potential and the potential
${\widetilde V}$ to be determined. The relevant Riccati equation to be
solved is now
\begin{equation}
\beta^2-\beta'= -g\delta(x)- \sigma,
\label{riccdelta}
\end{equation}
where $\sigma$ is the factorization energy in this case.  Remark that from
the mathematical point of view it is a differential equation which
includes a distribution. Therefore the solution could have some
discontinuity. It is possible to find a particular solution of
(\ref{riccdelta}), for a particular value of the constant $\sigma$, in
terms of the sign function, indeed $\beta_0(x)=(g/2)\, {\rm sgn\,}x$ for
$\sigma=-(g/2)^2$.  This function satisfies the differential equation
almost everywhere, {\it i.e.\/}, for every real value of $x$, except for
$x=0$. Then, the general solution can be found by using the standard
technique of transforming the nonlinear Riccati differential equation into
a linear one.  The final result is the following: 
\begin{equation}
\beta_\omega(x)=\frac{g}2\ {\rm sgn\,}x -\frac{d}{dx}\ln
\left(1- \omega \int_0^x e^{g|y|}\, dy\right)=
\frac{g}2\ {\rm sgn\,}x +\frac{ \omega\ e^{g|x|}}{1-\omega
\left( \frac{e^{g|x|}-1}g \right)  {\rm sgn\,}x}.
\label{susydelta}
\end{equation}
But this is precisely the result previously obtained in
(\ref{limitbetaminustodelta}) if we identify the two parameters
$\omega=\xi$. Obviously, the potential coming out from  this function will
be exactly the same as (\ref{limitvminustodelta}). Therefore, we have been
able to obtain also the general susy partners of the Dirac delta distribution
as a byproduct of the general results derived for the modified
P\"oschl-Teller potential in the previous section. We would like to insist on
the fact that in the singular potential case ($\xi\neq 0$) the susy
problem is related to a modification of the initial potential resulting from
adding an infinite barrier placed at the singularity (see \cite{Mar98}).

Let us comment now a little bit more on the results for the Dirac delta susy
partners. On one side, the particular case obtained from
(\ref{limitvminustodelta}) by making
$\xi=0$
\begin{equation}
\lim_{\alpha\to\infty} {\widetilde V}_0^- (\alpha,x)
=  +g\, \delta(x)=V^{\rm susy}_D(x)
\end{equation}
is in complete agreement with some already well known results
\cite{Boy88}. On the other side,  the general solution ($\xi\neq 0$)
presented in (\ref{limitvminustodelta}) can be compared with a result
recently published \cite{Boy98}. Some discrepancies can be appreciated
between the results derived in this last paper and ours. Observe that the
connection we have established between our results for the modified
P\"oschl-Teller and  Dirac delta potentials strongly supports the validity
of our new susy partners for the delta.
Finally, we would like to stress one of the main results: the
susy partners and all the relevant information for the Dirac delta potential
can be obtained taking the limit in the corresponding
expressions for the P\"oschl-Teller case, due to the good behaviour of the
limiting procedures. One remarkable difference is that for the delta potential
there is just one susy partner (indeed, only one specific value
of the factorization energy $\sigma$ allows to obtain the solution of the
Riccati equation (\ref{riccdelta})), in contrast to the analysis done for the
modified P\"oschl-Teller potentials, where two different factorization
energies $\epsilon^\pm$ were found.

\section{2--Susy modified P\"oschl-Teller potential}

The higher order susy partners can be also determined for the potentials 
we  considered before. 
Concerning this topic, it has been recently developed a handy technique
using difference equations in order to  construct multi-parametric families
of isospectral potentials \cite{Fer97,Ros98}. In this Section we shall
comment briefly only on the results derived following this approach for
the potentials we are dealing with in the 2--susy case. In the first susy
step we use the factorization constant
$\epsilon^+$ and in the second step we use $\epsilon^-$ (the same final
second order susy results are obtained if the process is accomplished in
the reverse order). Hence, the 2--susy potential is given by
\begin{equation}
V_{\zeta,\xi}^{+,-}(\alpha,x):= V(\alpha,x)-2\, \frac{d}{dx}\left(
\frac{\epsilon^+ - \epsilon^-}{\beta^+_\zeta (\alpha,x)- \beta^-_\xi
(\alpha,x)}  \right)
\end{equation}
where $V(\alpha,x)$ is the initial modified P\"oschl-Teller potential
(\ref{pt1param}), $\epsilon^\pm$ are given by (\ref{epsilonpm}), and
$\beta^+_\zeta(x)$, $\beta^-_\xi(x)$  by (\ref{betapluszeta}) and
(\ref{betaminusxi}), respectively. The 2--susy partner depends on the two
parameters $\zeta$ and $\xi$. This family embraces a wide variety of
potentials, one of the most interesting cases is obtained by taking only the
particular solutions $\zeta=\xi=0$:
\begin{equation}
V_{0,0} (\alpha,x)= -\frac{g \alpha}{2 \cosh^2 \alpha x}+
\frac{2 \alpha^2}{\sinh^2\alpha x}=
-\alpha^2\ \frac{\lambda
(\lambda-1)}{\cosh^2 \alpha x}+
\frac{2 \alpha^2}{\sinh^2\alpha x}.
\label{2susy}
\end{equation}
This kind of solution is a particular case of a more general form of
the modified P\"oschl-Teller potential used in some of the papers already
mentioned \cite{Bar87}
\begin{equation}
V(x)=\alpha^2\left( \frac{\kappa
(\kappa+1)}{\sinh^2 \alpha x}-\frac{\lambda
(\lambda-1)}{\cosh^2 \alpha x} 
\right),
\end{equation}
precisely for the value $\kappa=1$. Note that (\ref{2susy}) has not a well
defined limit when $\alpha\to\infty$, and therefore it is not possible to 
find a 2--susy partner for the Dirac delta connected with a 2--susy
partner of the modified P\"oschl-Teller potential. Observe that this fact
was implicit from the begining, because only one  factorization energy
$\sigma$ was found for the $\delta$, while for the modified
P\"oschl-Teller we were able to find two different factorization energies
$\epsilon^+$, $\epsilon^-$.

\section{Final remarks}

Due to the relevance of solvable susy quantum mechanical models as toy
examples for higher dimensional quantum field theories, and also for their
use in solid state physics, we analized in detail the 
supersymmetry associated with the modified  P\"oschl-Teller potential,
which appears in  many interesting physical situations, for example in the
nonrelativistic limit of the sine-Gordon equation, in connection with a
two-body force of Dirac delta type, when studying integrable many-body
systems in one dimension, or when considering two-dimensional susy
quantum field theories. 

In the present work we have constructed two new one-parametric 
families of exactly solvable potentials ${\widetilde V}_\zeta^+ (\alpha,x)$
and ${\widetilde V}_\xi^- (\alpha,x)$ related to the modified 
P\"oschl-Teller potential $V(\alpha,x)$ by the one-parametric
superpotentials
${\beta}_\zeta^+ (\alpha,x)$ and ${\beta}_\xi^- (\alpha,x)$, respectively. 
They represent two different cases of unbroken supersymmetry, and they
reduce to some previously published results for
$\zeta=0$ or $\xi=0$. A relevant trait of our results is that, for
specific values of $\zeta$ and $\alpha$, the members of the family
${\widetilde V}_\zeta^+ (\alpha,x)$  are free of singularities, which is, as 
far as we know, a fact unnoticed in the literature. On the other hand, the
family
${\widetilde V}_\xi^- (\alpha,x)$, with $\xi\neq 0$, embraces only 
singular potentials which have to be considered very carefully, because
they are not susy partners of the initial potential (\ref{pt1param}), but of
the initial potential plus an infinite barrier at the singularity \cite{Mar98}.

The connection between the modified  P\"oschl-Teller and Dirac delta 
potentials was established, and we were able to construct  the susy 
partner  of the delta potential. The main remark is that only the singular
family 
${\widetilde V}_\xi^- (\alpha,x)$ can be used to approximate an atractive
delta potential in terms of the limiting procedure discussed in the paper. 
The explanation can be given in terms of the susy process: our particular
solution of the Riccati equation (\ref{riccdelta}) represents the
superpotential
$\beta_0(x)=(g/2)\, {\rm sgn\,} x$ usually derived for the atractive delta
potential \cite{Boy88}. Therefore, the susy partner potential of the
Dirac delta well is a Dirac delta barrier and, because this last potential has
not bound states, the corresponding susy system has to present unbroken
susy; in other words, the susy process eliminates the only bound state of
the delta well in order to satisfy the Witten index condition for unbroken
susy. The same holds for the case $\omega=\xi\neq 0$. The energy level at
$\epsilon^-=E_0$ is destroyed and it does not play any role in the limit
$\alpha\to\infty$ for the spectrum of ${\widetilde V}_\xi^- (\alpha,x)$
(remember that $\lim_{\alpha\to\infty}E_0=E_\delta$). As the other 
energy levels dissapear, from the spectrum of ${\widetilde V}_\xi^-
(\alpha,x)$ when taking this limit, then the final potential has not bound
states. The result is in complete agreement with the direct calculation.
For ${\widetilde V}_\zeta^+ (\alpha,x)$, the situation is different 
because the ground energy level $\epsilon^+$ diverges as $-\alpha^2$
when $\alpha\to\infty$, and the potential has not a physically interesting
limit.

\section{Acknowledgements}
This work has been partially supported by a DGES project
(PB94--1115)  from Ministerio de Educaci\'on y Cultura (Spain), 
and also by Junta de Castilla y Le\'on (CO2/197). ORO acknowledges
support by CONACyT (Mexico), and the kind hospitality at the
Departamento de F\'{\i}sica Te\'orica (Univ. de Valladolid).

\newpage

\begin{figure}[htbp]
\centerline{
\epsfxsize=10truecm
\epsfbox{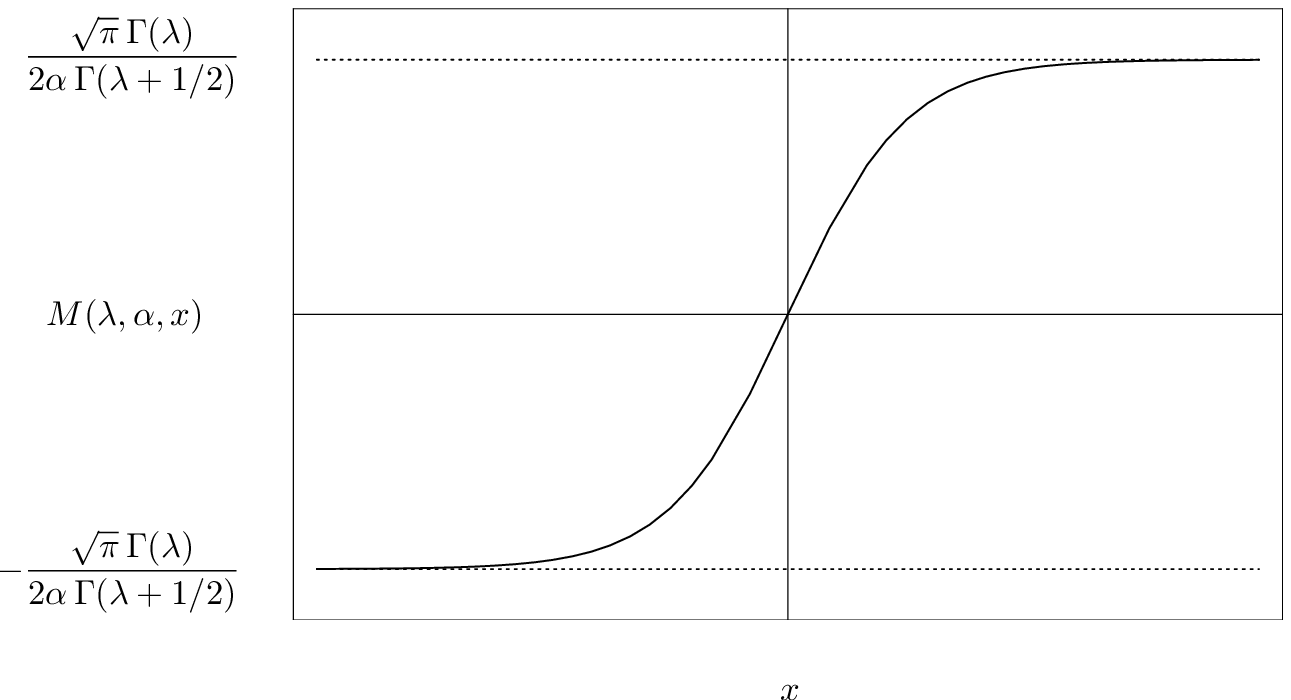}}
\begin{center}
\begin{minipage}{10cm}
\caption{
The function $M(\lambda,\alpha,x)$ given in Eq. (\ref{29}).  Its bounded
character enables the existence of non-singular susy partners ${\widetilde
V}^+_\zeta (\alpha,x)$. 
}
\end{minipage}
\end{center}  
\end{figure}
\begin{figure}[htbp]
\centerline{
\epsfxsize=10truecm
\epsfbox{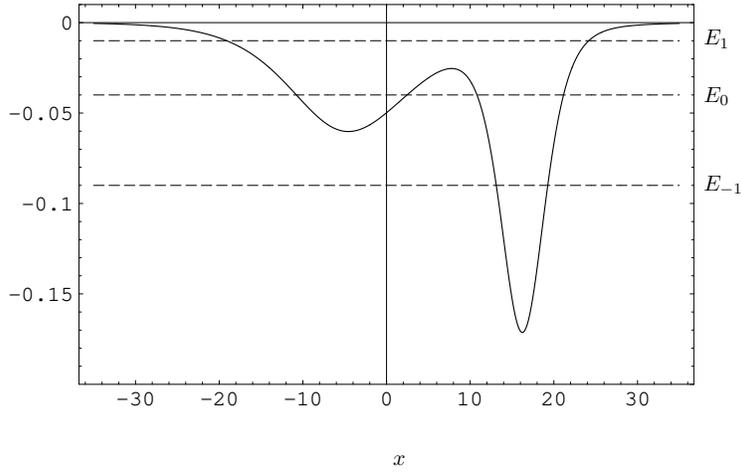}}
\begin{center}
\begin{minipage}{10cm}
\caption{
The susy partner potential ${\widetilde V}^+_\zeta (\alpha,x)$ given by
Eq. (\ref{vtildepluszeta}), with $\alpha=0.1$, $\lambda=3$, and
$\zeta=0.0937$, is an asymmetric double well. Its three bound states are
also represented with dotted horizontal lines. 
}
\end{minipage}
\end{center}  
\end{figure}
\begin{figure}[htbp]
\centerline{
\epsfxsize=10truecm
\epsfbox{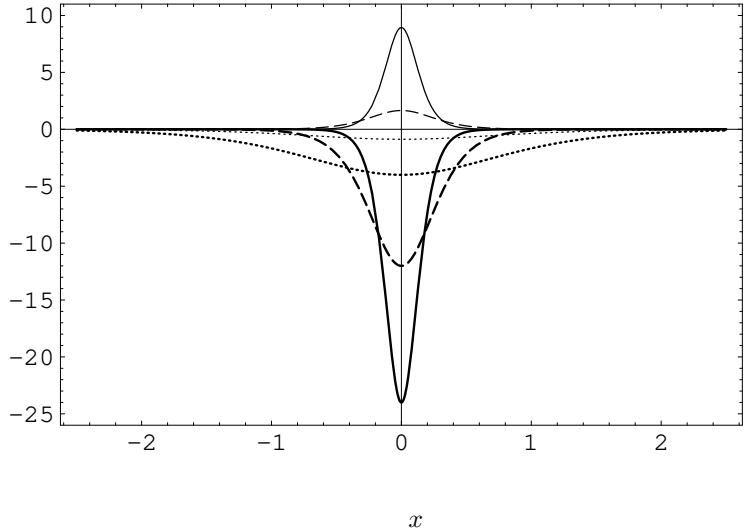}}
\begin{center}
\begin{minipage}{10cm}
\caption{
Different P\"oschl-Teller potentials $V(\alpha,x)$ from Eq.
(\ref{pt1param}) (the three thicker curves), and their corresponding susy
partner potentials ${\widetilde V}^-_0 (\alpha,x)$ given by Eq.
(\ref{vtildeminuszeta}) (the three thiner curves). The values of the
parameters are the following: dotted curves $\alpha=1$, $\lambda=2.562$,
dashed curves $\alpha=3$, $\lambda=1.758$, and solid curves $\alpha=6$,
$\lambda=1.457$. In all the cases the parameter $\xi$ is taken to be zero.
The initial P\"oschl-Teller potentials are always negative;  their susy
partners are less negative (dotted curve) or become even positive (dashed
and solid curves). 
}
\end{minipage}
\end{center}  
\end{figure}
\begin{figure}[htbp]
\centerline{
\epsfxsize=10truecm
\epsfbox{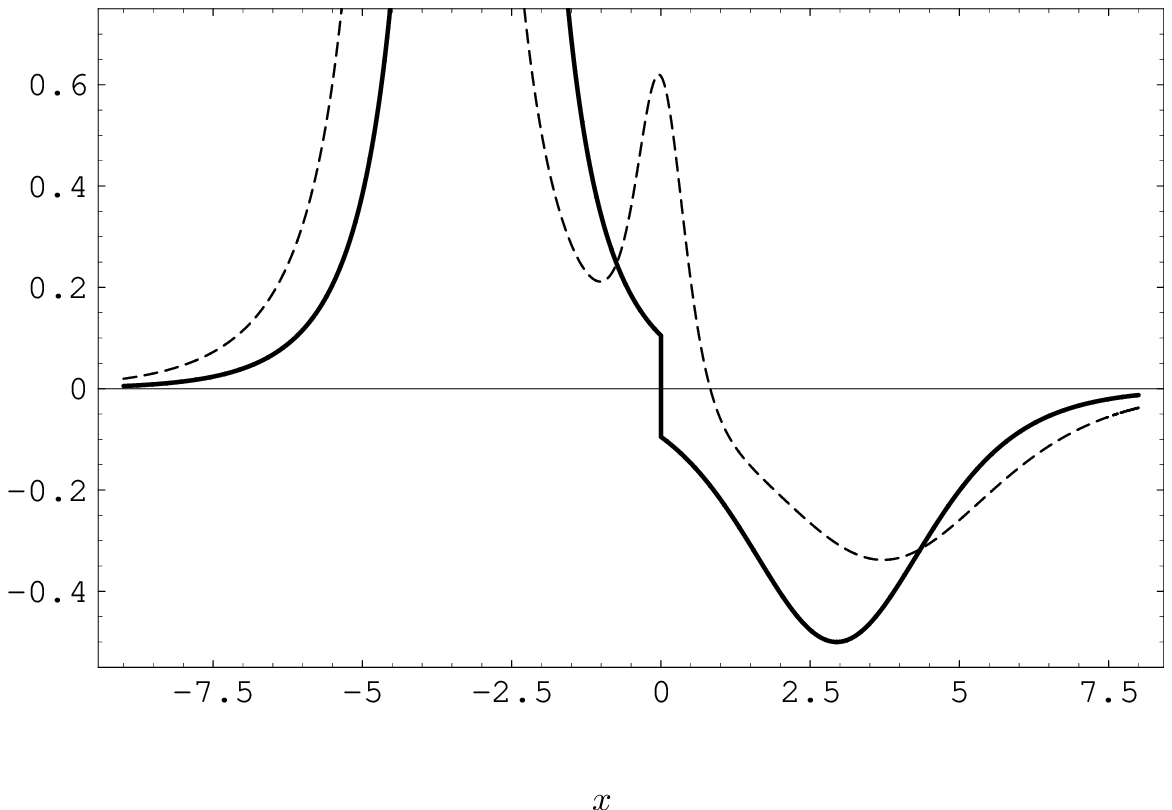}}
\begin{center}
\begin{minipage}{10cm}
\caption{
The dotted curve shows the plot of a member of the family of susy partner
P\"oschl-Teller potentials ${\widetilde V}^-_\xi (\alpha,x)$ given in Eq.
(\ref{vtildeminuszeta}), for the following values of the parameters: 
$\alpha=1.9$, $\lambda=1.216$ (or equivalently $g=1$), and $\xi=-0.05$.
The solid curve represents the susy partner Dirac delta potential for
$g=1$ and $\xi=-0.05$, and it is obtained from the dotted curve when the
limit $\alpha\to\infty$ is considered. In the same plot there are two
remarkable details: first the divergent term $g\,\delta(x)$ (generated by
the dotted hump) has not been represented, and second the potential has a
discontinuity at $x=0$, which is masked by the existence of the term
$g\,\delta(x)$. 
}
\end{minipage}
\end{center}  
\end{figure}


\begin{thebibliography}{99}

\bibitem{Ros32}
N. Rosen and P.M. Morse, Phys. Rev. {\bf 42}, 210 (1932).

\bibitem{Pos33}
G. P\"oschl and E. Teller, Z. Physik {\bf 83}, 143 (1933); W. Lotmar
Z. Physik {\bf 93}, 528 (1935).

\bibitem{Alh83} 
Y. Alhassid, F. G\"ursey, and F. Iachello, Phys. Rev. Lett. {\bf 50}, 873 (1983).

\bibitem{Nie78}
M.M.~Nieto, Phys. Rev. A {\bf 17}, 1273 (1978).

\bibitem{Bar87}
A.O.~Barut, A.~Inomata and R.~Wilson, J. Phys. A {\bf 20},
4083 (1987).

\bibitem{Inf51} 
L. Infeld and T.E. Hull, Rev. Mod. Phys. {\bf 23}, 21 (1951). 

\bibitem{Mat91}
V.B.~Matveev and M.A.~Salle, {\em Darboux Transformations and
Solitons}, Springer-Verlag, Berlin (1991).

\bibitem{Wit81}
E. Witten, Nucl. Phys. B {\bf 188}, 513 (1981).

\bibitem{Boy88}
L.J. Boya, Eur. J. Phys. {\bf 9}, 139 (1988).

\bibitem{Bra85}
H.W. Braden and A.J. Macfarlane, J. Phys. A {\bf 18}, 3151 (1985);
G.~Dunne and J. Feinberg, Phys. Rev. D {\bf 57}, 1271 (1998).

\bibitem{Mit98}
I.~Mitra, A.~Dasgupta and B.~Dutta-Roy, Am. J. Phys. {\bf 66}, 1101 (1998).

\bibitem{Atk75}
D.A.~Atkinson and H.W.~Crater, Am. J. Phys. {\bf 43}, 1301 (1975).

\bibitem{Ols83}
M.A.~Olshanetsky and A.M~Perelomov, Phys. Rep. {\bf 94}, 313 (1983).

\bibitem{Dar82} 
G. Darboux, C. R. Acad. Sci. Paris {\bf 94}, 1456 (1882). 

\bibitem{Mie84}
B. Mielnik, J. Math. Phys. {\bf 25}, 3387 (1984).

\bibitem{Fer84}
D.J.~Fern\'andez C., Lett. Math. Phys. {\bf 8}, 337 (1984) .

\bibitem{Suk85}
C.V. Sukumar, {\em J. Phys. A} {\bf 19}, 2297 (1986)

\bibitem{Zak90}
B.N.~Zakhariev and A.A.~Suzko, {\em Direct and Inverse Problems},
Springer-Verlag, New York (1990).

\bibitem{Car79}
R.W.~Carrol, {\em Transmutation and Operator Differential equations} vol
37, North-Holland, Amsterdan (1979).

\bibitem{Fer97}
D.J.~Fern\'andez C., M.L.~Glasser and L.M.~Nieto, Phys. Letts A {\bf 240}, 15
(1998); D.J.~Fern\'andez C., V.~Hussin and B.~Mielnik, Phys. Lett. A {\bf
244}, 309 (1998).

\bibitem{Ros98}
J.O.~Rosas-Ortiz, { J. Phys. A} {\bf 31}, L507 (1998);
{ J. Phys. A} {\bf 31}, 10163 (1998).

\bibitem{And91}
A.~Anderson and R.H.~Price, Phys. Rev. D {\bf 43}, 3147 (1991). 

\bibitem{Mar98}
I.F. M\'arquez, J. Negro, and L.M. Nieto,  J.~Phys.~A {\bf 31}, 
4115 (1998).

\bibitem{Lan65}
L.D. Landau and E.M. Lifshitz, {\em Quantum Mechanics}, Pergamon (1965).
S. Fl\"ugge, {\em Practical Quantum Mechanics}, Springer-Verlag, New York
(1974).

\bibitem{Erd53} 
A. Erd\'elyi, {\em Higher Transcendental Functions, Vol. I}, McGraw-Hill
(1953). 

\bibitem{Boy98}
L.J. Boya, H.C. Rosu, A.J. Segu\'{\i}-Santonja, J.  Socorro, F.J. Vila, J.
Phys. A {\bf 31}, 8835 (1998).


\end{thebibliography}
\end{document}